\documentclass[12pt]{article}
\usepackage{amsmath,amssymb,amsfonts,graphicx}

\begin{document}
\thispagestyle{empty}
\renewcommand{\refname}{References}

\title{\bf Scattering theory and \\ the Aharonov--Bohm effect in \\ quasiclassical physics}
\author{Yurii A.~Sitenko$^{1}$ and Nadiia D.~Vlasii$^{1,2}$}
\date{}
\maketitle
\begin{center}
$^{1}$Bogolyubov Institute for Theoretical Physics, \\
National Academy of Sciences, 03680, Kyiv, Ukraine \\
$^{2}$Physics Department, Taras Shevchenko National University of
Kyiv, \\
01601, Kyiv, Ukraine
\end{center}
\medskip
\medskip

Scattering of a nonrelativistic quantum-mechanical particle by an
impenetrable magnetic vortex is considered. The nonvanishing
transverse size of the vortex is taken into account, and the limit
of short, as compared to this size, wavelengths of the scattered
particle is analyzed. We show that the scattering Aharonov-Bohm
effect persists in the quasiclassical limit owing to the diffraction
persisting in the short-wavelength limit. As a result, the vortex
flux serves as a gate for the propagation of short-wavelength,
almost classical, particles. This quasiclassical effect is more
feasible to experimental detection in the case when space outside
the vortex is conical.


\begin{center}
\end{center}

\medskip
\medskip

\section{Introduction}

The Aharonov-Bohm effect \cite{Aha} (somewhat partly anticipated
earlier by Ehrenberg and Siday \cite{Ehr}) plays a fundamental role
in modern physics. It demonstrates that quantum matter is influenced
by electromagnetic field even in the case when the region of
nonvanishing field strength does not overlap with the region
accessible to quantum matter; the indispensable condition is that
the latter region be non-simply-connected. A particular example is a
magnetic field of an infinitely long solenoid which is shielded and
made impenetrable to quantum matter; such a field configuration may
be denoted as an impenetrable magnetic vortex. Although nowadays the
Aharonov-Bohm effect is generalized in various aspects and in
different areas of modern physics, in the present paper we shall
discuss its traditional formulation as of a quantum-mechanical
scattering effect off an impenetrable magnetic vortex (see reviews
in \cite{Rui,Ola,Pes,Afa,Ton}). Even in this restricted sense, it
corresponds to two somewhat different but closely related setups.
The first one concerns the fringe shift in the interference pattern
due to two coherent particle beams under the influence of an
impenetrable magnetic vortex placed between the beams. The second
one deals with scattering of a particle beam directly on an
impenetrable magnetic vortex. Almost all experiments are performed
in the first setup, though the second setup is more elaborate from
the theoretical point of view. A direct scattering experiment
involving long-wavelength (slow-moving) particles is hardly
possible, but that involving short-wavelength (fast-moving)
particles is quite feasible, and in the present paper we propose to
perform such an experiment.

Since the Aharonov-Bohm effect is the purely quantum effect which
has no analogues in classical physics \footnote{ It should be noted
in this respect that, on the contrary, the geometric phase of Berry
\cite{Be1} has its classical analogue~ --~ Hannay's angle
\cite{Han,Be2}.}, it becomes evidently more manifest in the limit of
long wavelengths of a scattered particle, when the wave aspects of
matter are exposed to the maximal extent. As the particle wavelength
decreases, the wave aspects of matter are suppressed in favour of
the corpuscular ones, and therefore the persistence of the
Aharonov-Bohm effect in the limit of short (as compared to the
vortex thickness) wavelengths seems to be rather questionable.
Actually, there is a controversy in the literature concerning this
point. As it follows from \cite{Rui,Ola}, scattering off an
impenetrable magnetic vortex in the short-wavelength limit tends to
classical scattering off a tube of hard core, which is independent
of the enclosed magnetic flux, and thus the Aharonov-Bohm effect is
extinct in this limit. On the other hand, it was already shown by
Aharonov and Bohm \cite{Aha} for the case of an idealized
(infinitely thin) vortex that the wave function vanishes in the
strictly forward direction, when the vortex flux equals a
half-of-odd-integer multiple of the London flux quantum; later this
result was generalized to the case of a realistic vortex of finite
thickness \cite{Afa} and, being independent of the value of the
particle wavelength, it persists in the short-wavelength limit.
Thus, this circumstance witnesses in favour of the persistence of
the Aharonov-Bohm effect, since the wave function for all other
values of the vortex flux is for sure nonvanishing in the forward
direction.

The exclusiveness of the forward direction is of no surprise. We
recall the well-known fact that the short-wavelength limit of
quantum-mechanical scattering off a hard core does not converge with
the classical point-particle scattering (perfect reflection), it
differs by a forward peak which is due to the Fraunh\"{o}fer
diffraction; the peak is increasing, as the particle wavelength is
decreasing and the particle is becoming like a classical point
corpuscle. Meanwhile the width of the forward peak is decreasing,
and that is why the experimental detection of the peak is a rather
hard task. As is noted in \cite{Mor}, it seems more likely that the
measurable quantity is the classical cross section, although the
details of this phenomenon depend on the method of measurement. On
the other hand, the forward peak cannot in any way be simply
ignored, because its amplitude is involved in the optical theorem,
whereas the amplitude yielding classical scattering vanishes in the
forward direction. Also, the forward peak contributes considerably
to the total cross section, making the latter twice as large as the
classical total cross section.

Thus, the quantum-mechanical scattering effects persist in the
quasiclassical limit owing to the diffraction effects persisting in
the short-wavelength limit. Concerning the Aharonov-Bohm effect,
this conjecture is justified quantitatively in the present paper.

However, our consideration is focused mainly on the scattering
Aharonov-Bohm effect in conical space. Conical space is a space
which is locally flat almost everywhere with exception of a region
in the form of an infinitely long tube; the metric outside the tube
is given by squared length element \cite{Mar, Isr, Sok}
\begin{equation}
{\rm d}s^2=(1-\eta)^{-2}{\rm d}\tilde{r}^2+\tilde{r}^2{\rm d}\varphi^2+{\rm d}z^2={\rm d}r^2+r^2{\rm d}\tilde{\varphi}^2
+{\rm d}z^2,\label{eq1}
\end{equation}
where
$$
\tilde{r}=r(1-\eta),\qquad 0<\varphi<2\pi, \qquad 0<\tilde{\varphi}<2\pi(1-\eta),
$$
and $\eta$ is related to the curvature integrated over the
transverse section of the tube, being of the same sign. Deficit
angle $2\pi\eta$ is bounded from above by $2\pi$ and is unbounded
from below (quantity $-2\pi\eta$ for negative $\eta$ is the proficit
angle that can be arbitrarily large), thus $-\infty<\eta<1$. Conical
space emerges inevitably as an outer space of a topological defect
in the form of a string; such defects known as the
Abrikosov-Nielsen-Olesen vortices \cite{Abr, Nie} arise as a
consequence of phase transitions with spontaneous breakdown of gauge
symmetries, when the first homotopy group of the group space of the
broken symmetry group is nontrivial. Certainly, the value of $\eta$
is vanishingly small for vortices in superconductors, but vortices
under the name of cosmic strings \cite{Kib, Vil} are currently
discussed in cosmology and astrophysics, and the observational data
is consistent with the values of $\eta$ in the range $0<\eta<4\cdot
10^{-7}$ (see, e.g., \cite{Bat}), although the direct evidence for
the existence of cosmic strings is still lacking. In carbon
nanophysics, topological defects in graphene (two-dimensional
crystal of carbon atoms) correspond to nanocones with the values of
$\eta$ equal to positive and negative multiples of $1/6$
\cite{Si7,Si8}. At last, conical space may emerge in a rather
general context of contemporary condensed matter physics which
operates with a variety of two-dimensional structures (thin films)
made of different materials. If such a film is rolled into a cone,
then one can generate quasiparticle excitations in this
conically-shaped film and consider their propagation towards and
through the tip. In all above setups, the problem of
quantum-mechanical scattering of a nonrelativistic particle by a
magnetic vortex in conical space may be relevant.

Scattering in an idealized (with the core of zero transverse size)
conical space was considered by 't Hooft \cite{Ho} and Jackiw et al
\cite{Des,Sou}; later the consideration was extended to the case of
an idealized magnetic vortex placed along the axis of an idealized
conical space \cite{Si2}. However, in the quasiclassical limit the
effects of nonzero transverse size of the core become significant.
These effects were taken properly into account in \cite{Si5} (see
also \cite{SiV,S10}), and we shall implicate the results of the
latter works. Instead of using the quasiclassical WKB approximation
or an analogue of the Kirchhoff approximation in optics, we shall
get the quasiclassical limit directly from exact expressions for the
scattering amplitude.

In the next section we consider scattering of a classical point
particle by an impenetrable tube of hard core in conical space. In
Section 3 we consider scattering of a quantum-mechanical
nonrelativistic particle by an impenetrable tube in conical space;
the quasiclassical limit is obtained. In Section 4 we solve the same
problem for the case when the tube is filled with the magnetic flux
lines, i.e. for the case of an impenetrable magnetic vortex. The
results are discussed and summarized in Section 5, the conclusions
are drawn in Section 6. Some basic provisions of quantum-mechanical
scattering theory in conical space are reviewed in Appendix.

\section{Classical scattering off an impenetrable tube in conical space}

Scattering of a classical point particle off an impenetrable tube in
Euclidean space is a well-known simple problem, the solution to
which can be found in a textbook on classical mechanics (see, e.g.,
\cite{Lan}). The incidence angle is equal to the reflection angle
and is denoted in the following by $\chi$. We consider the case when
incident particles fall perpendicularly to the axis of the tube of
radius $r_c$. Scattering angle $\varphi$ is related to angle $\chi$,
$\varphi=\pi-2\chi$, and the latter defines impact parameter
$\rho=r_c\sin\chi$, see Fig. 1 where the tube is directed
perpendicular to the plane of the figure and its position is
indicated by the circle. When space outside the tube is conical, the
relation between the angles is modified:
\begin{equation}
(1-\eta)(\varphi-\varphi_\eta)=\pi-2\chi, \label{eq2}
\end{equation}
where $2\pi\eta$ is the deficit angle of conical space, see (1), and
$\varphi_\eta$ is connected with the purely kinematic scattering
angle for particles with impact parameters exceeding the transverse
size of the tube. It is instructive to consider cases of both
positive and negative values of the deficit angle here. In the case
$-\infty>\eta>0$, the trajectories of particles bypassing the tube
without an impact diverge, and the region of angles
$-\omega_\eta<\varphi<\omega_\eta$ (where
$\omega_\eta=-\eta\pi(1-\eta)^{-1}$) can be denoted as the region of
shadow, see Fig. 2. In the case $0<\eta<1/2$, the trajectories of
particles bypassing the tube without an impact converge (and
intersect), and the region of angles
$-\omega_\eta<\varphi<\omega_\eta$ (where
$\omega_\eta=\eta\pi(1-\eta)^{-1}$) can be denoted as the region of
double image, see Fig. 3. The maximal value of angle $\chi$, which
is $\pi/2$, corresponds to the minimal value of angle $\varphi$,
which is $\varphi_\eta$, and the latter is either positive (in space
with the shadow region), or negative (in space with the double-image
region):
\begin{equation}
\varphi_\eta=-\eta\pi(1-\eta)^{-1}. \label{eq3}
\end{equation}
Thus, the impact parameter for particles which undergo scattering by
the tube is
\begin{equation}
\rho=r_c(1-\eta)\sin\chi=r_c(1-\eta)\sin\left[\frac 12(1-\eta)(\pi-\varphi)\right], \label{eq4}
\end{equation}
where the appearance of factor $(1-\eta)$ after $r_c$ is due to the
fact that the circumference of the tube in conical space is $2\pi
r_c(1-\eta)$.

As in the case of Euclidean space ($\eta=0$), the differential cross
section is related to the derivative of the impact parameter on the
scattering angle, and we get the differential cross section for
incident particles in the upper half-space (with respect to the
horizontal plane passing through the axis of the tube):
\begin{equation}
\frac{{\rm d}\sigma}{{\rm d}\varphi}=-\frac{{\rm d}\rho}{{\rm d}\varphi}=\frac 12r_c(1-\eta)^2
\sin\left[\frac 12(1-\eta)\varphi+\frac 12\eta\pi\right],\quad
\varphi_\eta<\varphi<\pi. \label{eq5}
\end{equation}
A similar consideration yields the differential cross section for
incident particles in the lower half-space:
\begin{equation}
\frac{{\rm d}\sigma}{{\rm d}\varphi}=\frac{{\rm d}\rho}{{\rm d}\varphi}=\frac 12r_c(1-\eta)^2
\sin\left[\frac 12(1-\eta)\varphi+\frac 12\eta\pi\right],\quad
\pi<\varphi<2\pi-\varphi_\eta. \label{eq6}
\end{equation}

It should be emphasized that, although expressions (5) and (6) are
true for $-\infty<\eta<1/2$, their physical consequences for
negative and positive values of the deficit angle are quite
different. Since the shadow region cannot be reached by incident
particles, the total cross section in the case $-\infty<\eta<0$ is
\begin{equation}
\sigma=\int\limits_{\omega_\eta}^{2\pi-\omega_\eta}{\rm d}\varphi\frac{{\rm d}\sigma}{{\rm d}\varphi}=2r_c(1-\eta)
\quad (\omega_\eta=\varphi_\eta). \label{eq7}
\end{equation}
On the contrary, the double-image region can be reached by incident
particles from both upper and lower half-spaces. Therefore, in the
case $0<\eta<1/2$ we get the cross section out of the double-image
region,
\begin{equation}
\sigma_{\rm out}=\int\limits_{\omega_\eta}^{2\pi-\omega_\eta}{\rm d}\varphi\frac{{\rm d}\sigma}{{\rm d}\varphi}=2r_c(1-\eta)\cos(\eta\pi)
\quad (\omega_\eta=-\varphi_\eta), \label{eq8}
\end{equation}
and the cross section in the double-image region which is reached
from either upper or lower half-space,
\begin{equation}
\sigma_{\rm in}=\int\limits_{-\omega_\eta}^{\omega_\eta}{\rm d}\varphi\frac{{\rm d}\sigma}{{\rm d}\varphi}=r_c(1-\eta)[1-\cos(\eta\pi)]
\quad (\omega_\eta=-\varphi_\eta). \label{eq9}
\end{equation}
The total cross section in this case is
\begin{equation}
\sigma=\sigma_{\rm out}+2\sigma_{\rm in}=2r_c(1-\eta). \label{eq10}
\end{equation}
Final expressions (7) and (10) for the total cross section allow one
to interpret them simply as the quotient of the circumference of the
tube to the half of the complete azimuthal angle (note that only
half of the tube is exposed to incident particles).

In conclusion of this section we note that the case $1/2<\eta<1$
which includes (in turn as $\eta$ increases) spaces with shadow and
double-image regions can be considered as well.

\section{Quasiclassical limit of quantum-mechanical scattering off an impenetrable tube in conical space}

In quantum mechanics, one considers a scattering wave solution to
the Schr\"{o}dinger equation ${\rm i}\hbar\partial_t\psi=H\psi$. In
the case of a cylindrically symmetric potential, the asymptotics of
this solution at large distances from the symmetry axis takes form
\begin{equation}
\psi\sim\exp(-{\rm i}Et\hbar^{-1}+{\rm i}k_zz)[\psi_{\rm in}({\bf r};\,{\bf k})+
f(k,\,\varphi)\exp({\rm i}kr)r^{-1/2}+O(r^{-1})], \label{eq11}
\end{equation}
where ${\bf k}$ is the two-dimensional wave vector which is
orthogonal to the symmetry axis, $k^2=2mE\hbar^{-2}-k^2_z>0$, $m$
and $E$ are the mass and the energy of a scattered particle. If the
potential corresponds to an impenetrable tube of radius $r_c$ and
infinite length, then the particle wave function obeys the Dirichlet
boundary condition at the edge of the tube
\begin{equation}
\psi|_{r=r_c}=0. \label{eq12}
\end{equation}
If space outside the tube is Euclidean, then the incident wave can
be a plane wave in, say, the $\varphi=0$ direction,
\begin{equation}
\psi_{\rm in}({\bf r};\,{\bf k})=\exp({\rm i}kr\cos \varphi), \label{eq13}
\end{equation}
and the scattering amplitude is given by expression
\begin{equation}
f(k,\,\varphi)=-\sqrt{\frac{2}{\pi {\rm i}k}}\sum\limits_{n\in \mathbb{Z}}e^{{\rm i}n\varphi}
\frac{J_{|n|}(kr_c)}{H_{|n|}^{(1)}(kr_c)}, \label{eq14}
\end{equation}
where $J_\nu(u)$ and $H_\nu^{(1)}(u)$ are the Bessel and the
first-kind Hankel functions of order $\nu$, $\mathbb{Z}$ is the set
of integer numbers.

The case of a fast-moving incident particle, $kr_c\gg 1$,
corresponds to the quasiclassical limit when the laws of geometric
(ray) optics seem to be applicable. One can get this limit directly
from exact expression (14) which is valid for all values of $k$. The
crucial point of the analysis is that the Bessel function becomes
vanishingly small when its order exceeds its large argument and,
therefore, the infinite sum in (14) is cut at $|n|\sim kr_c$ in the
case $kr_c\gg 1$ (see, e.g., \cite{New}). To be more precise, we
rewrite (14) as
\begin{equation}
f(k,\varphi)=-\frac{1}{\sqrt{2\pi {\rm i}k}}\left[
\sum\limits_{|n| \leq kr_c}e^{{\rm i}n\varphi}+
\sum\limits_{|n| \leq kr_c}e^{{\rm i}n\varphi}\frac{H_{|n|}^{(2)}(kr_c)}
{H_{|n|}^{(1)}(kr_c)}+2\sum\limits_{|n|>kr_c}e^{{\rm i}n\varphi}
\frac{J_{|n|}(kr_c)}{H_{|n|}^{(1)}(kr_c)}\right]. \label{eq15}
\end{equation}
where $H_\nu^{(2)}(u)$ is the second-kind Hankel function of order
$\nu$. The first sum in (15) is easily taken as a sum of geometric
progression, and its contribution to the scattering amplitude is
\begin{equation}
f^{({\rm peak})}(k,\,\varphi)=-\frac{1}{\sqrt{2\pi {\rm i}k}}\frac{\sin[([\![kr_c]\!] +1/2)\varphi]}
{\sin(\varphi/2)}, \label{eq16}
\end{equation}
where $[\![x]\!]$ denotes the integer part of quantity $x$ (i.e. the
integer which is less than or equal to $x$). In the case $kr_c\gg
1$, when $[\![kr_c]\!]+1/2\approx kr_c$, (16) is strongly peaked in
the strictly forward direction, $\varphi=0$, vanishing otherwise:
\begin{equation}
f^{(\rm peak)}(k,\,\varphi)=-\sqrt{\frac{2k}{\pi {\rm i}}}\,r_c\{1+O[( kr_c\varphi)^2]\},\quad |\varphi|\ll(kr_c)^{-1}. \label{eq17}
\end{equation}
In view of relation
$$
\frac{1}{4\pi x}\int\limits_{-\pi}^{\pi}d\varphi\,\frac{\sin^2(x\varphi)}{\sin^2(\varphi/2)}=1+O(x^{-2}),\,\,\,\,x\gg 1,
$$
one can introduce function
\begin{equation}
\Delta_x(\varphi)=\frac{1}{4\pi x}\frac{\sin^2(x\varphi)}{\sin^2(\varphi/2)}\,\,\,\,
(-\pi<\varphi<\pi), \label{eq18}
\end{equation}
as a regularized (smoothed) delta-function for the angular variable:
\begin{equation}
\lim\limits_{x\rightarrow\infty}\Delta_x(\varphi)=\Delta(\varphi)\equiv\frac1{2\pi}\sum\limits_{n\in\mathbb{Z}}
e^{{\rm i}n\varphi}, \quad \Delta_x(0)=\frac x\pi. \label{eq19}
\end{equation}
Then (16) in the case $kr_c\gg 1$ is rewritten as
\begin{equation}
f^{(\rm peak)}(k,\varphi)=-\sqrt{-2{\rm i}r_c\,\Delta_{kr_c}(\varphi)}. \label{eq20}
\end{equation}
One can find the appropriate differential cross section,
\begin{equation}
\frac{{\rm d}\sigma^{(\rm peak)}}{{\rm d}\varphi}=\left|f^{(\rm peak)}(k,\,\varphi)\right|^2
=2r_c\Delta_{kr_c}(\varphi), \label{eq21}
\end{equation}
and the appropriate integrated cross section,
\begin{equation}
\sigma^{(\rm peak)}=\int\limits_{-\pi}^{\pi}{\rm d}\varphi\frac{{\rm d}\sigma^{(\rm peak)}}{{\rm d}\varphi}=2r_c. \label{eq22}
\end{equation}
The forward peak exhibited by (21) is due to the Fraunh\"{o}fer
diffraction that is the diffraction in almost parallel rays. It
should be noted that, since the angular delta-function can be
regularized in different ways, the expressions for $f^{(\rm
peak)}(k,\,\varphi)$ can differ from (16) (see \cite{New,Mor}), but
all of them correspond to $\Delta_{kr_c}(\varphi)$ with properties
(19).

The second sum in (15) requires a more elaborate analysis involving
the use of the stationary phase method, and its contribution to the
scattering amplitude in the case $kr_c\gg 1$ is
\begin{equation}
f^{(\rm class)}(k,\,\varphi)=-\sqrt{\frac{r_c}{2}\sin\frac{\varphi}{2}}\exp(-2{\rm i}kr_c\sin\frac{\varphi}{2})+
\sqrt{r_c}O[(kr_c)^{-1/2}]\,\,\,\, (0<\varphi<2\pi). \label{eq23}
\end{equation}
Unlike (20), this contribution is rather uniform in the scattering
angle, smoothly vanishing in the forward direction. The appropriate
differential cross section,
\begin{equation}
\frac{{\rm d}\sigma^{(\rm class)}}{{\rm d}\varphi}=|f^{(\rm class)}(k,\,\varphi)|^2=\frac{r_c}{2}\sin\frac \varphi 2, \label{eq24}
\end{equation}
describes classical scattering (elastic reflection) of a point
particle. The appropriate integrated cross section,
\begin{equation}
\sigma^{(\rm class)}=\int\limits_{0}^{2\pi}{\rm d}\varphi\frac{{\rm d}\sigma^{(\rm class)}}{{\rm d}\varphi}=2r_c, \label{eq25}
\end{equation}
coincides in its value with (22). The contribution of the third sum
in (15) is estimated to be of the order
$\sqrt{r_c}O[(kr_c)^{-1/6}]$, and it can be neglected in the case
$kr_c\gg 1$.

Thus, the quasiclassical limit does not coincide with classical
theory: geometric optics is supplemented by the Fraunh\"{o}fer
diffraction. The incident wave in the process of scattering is
transformed into two parts: the one is reflected in all directions
according to the laws of geometric optics, and the other one forms
the diffraction peak in the forward direction. Each part yields a
cross section equal to the diameter of the tube, see (25) and (22),
and the total cross section,
\begin{equation}
\sigma_{\rm tot}=\sigma^{(\rm class)}+\sigma^{(\rm peak)}=4r_c, \label{eq26}
\end{equation}
is twice the classical total cross section. Namely the total cross
section (as well as $f^{({\rm peak})}(k,\,0)$, since $f^{({\rm
class})}(k,\,0)=0$) is involved in the optical theorem which
expresses the probability conservation,
\begin{equation}
2\sqrt{\frac{2\pi}{k}}{\rm Im}\left[{\rm i}^{-1/2}f(k,\,0)\right]=\sigma_{\rm tot}. \label{eq27}
\end{equation}
However, due to the extremely narrow width of the diffraction peak,
it is rather hard to detect this peak experimentally \cite{Mor}.

If space outside the tube is conical, then the exact expression for
the scattering amplitude is
\begin{eqnarray}
f(k,\,\varphi)&=&-\exp[2{\rm i}k(r_c-\xi_c)]\sqrt{\frac{2}{\pi {\rm i}k}}\sum\limits_{n\in \mathbb{Z}}e^{{\rm i}n(\varphi-\pi)}
\left[\frac {\rm i}2\sin\left(\frac{|n|\pi}{1-\eta}\right)+\right. \nonumber \\
&+&\left.\exp\left(-\frac{{\rm i}|n|\pi}{1-\eta}\right)\frac{J_{|n|(1-\eta)^{-1}}(kr_c)}
{H^{(1)}_{|n|(1-\eta)^{-1}}(kr_c)}\right], \label{eq28}
\end{eqnarray}
where $\xi_c=\int\limits_{0}^{r_c}{\rm d}s$ is the geodesic radius
of the tube (note that the spatial region inside the tube is
characterized by nonzero curvature). The contribution of the
infinite sum with the first (sine) term in square brackets yields
the amplitude for scattering on a tube of zero radius \cite{Ho,Des},
which is negligible in the quasiclassical limit. The remaining
infinite sum which contains the quotient of the cylindrical
functions is analyzed along the same lines as in the case of
Euclidean space. The dominant contribution is given by terms with
$|n| \leq kr_c(1-\eta)$ when $kr_c\gg 1$. The diffraction peaks are
now directed at $\varphi=\mp \varphi_\eta$ with $\varphi_\eta$ given
by (3), and the scattering amplitude in the quasiclassical limit
takes form
\begin{equation}
f(k,\,\varphi)=f^{({\rm peak})}_{+}(k,\,\varphi+\varphi_\eta)
+f^{(\rm peak)}_{-}(k,\,\varphi-\varphi_\eta)+
f^{(\rm q-class)}(k,\,\varphi)+\sqrt{r_c}O[(kr_c)^{-1/6}]. \label{eq29}
\end{equation}
Here the contribution of the Fraunh\"{o}fer diffraction on the tube
is given by
\begin{eqnarray}
f_{\pm}^{({\rm peak})}(k,\,\varphi\pm \varphi_\eta)&=&-\frac{\exp
\left[2{\rm i}k(r_c-\xi_c)\right]}{\sqrt{2\pi {\rm i}k}}\,\frac{\sin\left[\frac 12kr_c(1-\eta)(\varphi\pm\varphi_\eta)\right]}{\sin\left[\frac 12
(\varphi\pm\varphi_\eta)\right]}\times\nonumber \\
&\times&
\exp\left\{\frac{{\rm i}}{2}\left[
1\pm kr_c(1-\eta)\right](\varphi\pm\varphi_\eta)\right\},
\label{eq30}
\end{eqnarray}
and the contribution of the quasiclassical reflection from the tube
is given by (see \cite{Si5})
\begin{eqnarray}
f^{({\rm q-class})}(k,\,\varphi)=-\exp[2{\rm i}k(r_c-\xi_c)](1-\eta)\sqrt{\frac{r_c}{2}}\times \nonumber \\
\times\sum\limits_{l}\sqrt{\cos\left[\frac 12(1-\eta)(\varphi-\pi+2l\pi)\right]}
\exp\left\{-2{\rm i}kr_c\cos\left[\frac 12(1-\eta)(\varphi-\pi+2l\pi)\right]\right\}, \label{eq31}
\end{eqnarray}
where the finite sum is over integers $l$ satisfying condition
\begin{equation}
-(2\pi)^{-1}(\varphi-\varphi_\eta)<l<-(2\pi)^{-1}(\varphi+\varphi_\eta)+1. \label{eq32}
\end{equation}
It should be noted that the incident wave in this case differs from
(13) taking the form \cite{Ho,Des}
\begin{equation}
\psi_{{\rm in}}({\bf r};\,{\bf k})=(1-\eta)\sum\limits_{l}\exp\{-{\rm i}kr\,\cos[(1-\eta)
(\varphi-\pi+2l\pi)]\}, \label{eq33}
\end{equation}
where the summation is over the same values of $l$ as given by (32).

The number of terms in (31) or (33), which is defined by (32) and is
denoted in the following by $n_l$, depends on, whether the values of
the scattering angle are in the range out or in the classical
shadow, see Fig. 2, or in the range out or in the classical double
image, see Fig. 3. The value of $n_l$ outside the shadow region is
odd and larger by 1 than that inside the shadow region, while the
value of $n_l$ outside the double-image region is also odd but
smaller by 1 than that inside the double-image region. In the case
$-\infty<\eta<0$ we have $n_l=1$ outside and $n_l=0$ inside, and the
differential cross section corresponding to (31) is
\begin{equation}
\frac{{\rm d}\sigma^{({\rm q-class})}}{{\rm d}\varphi}=\frac 12r_c(1-\eta)^2\sin\left[\frac 12(1-\eta)\varphi
+\frac 12\eta\pi\right],\quad \varphi_\eta<\varphi<2\pi-\varphi_\eta, \label{eq34}
\end{equation}
and
\begin{equation}
\frac{{\rm d}\sigma^{({\rm q-class})}}{{\rm d}\varphi}=0,\quad -\varphi_\eta<\varphi<\varphi_\eta. \label{eq35}
\end{equation}
In the case $0<\eta<1/2$ we have $n_l=1$ outside and $n_l=2$ inside,
and the differential cross section corresponding to (31) is
\begin{equation}
\frac{{\rm d}\sigma^{({\rm q-class})}}{{\rm d}\varphi}=\frac 12r_c(1-\eta)^2\sin\left[\frac 12(1-\eta)\varphi+
\frac 12\eta\pi\right],\quad -\varphi_\eta<\varphi<2\pi+\varphi_\eta, \label{eq36}
\end{equation}
and
\begin{eqnarray}
\frac{{\rm d}\sigma^{({\rm q-class})}}{{\rm d}\varphi}=r_c(1-\eta)^2\left\{\cos\left[\frac 12(1-\eta)\varphi\right]
\sin \left(\frac 12\eta\pi\right)+\right. \nonumber \\
\left.+\sqrt{\sin^2\left(\frac 12\eta\pi\right)-\sin^2\left(\frac 12(1-\eta)\varphi\right)}\,\cos\left[4kr_c
\sin\left(\frac 12(1-\eta)\varphi\right)\cos\left(\frac 12\eta\pi\right)\right]\right\},\nonumber \\
\varphi_\eta<\varphi<-\varphi_\eta. \label{eq37}
\end{eqnarray}
One can note the consistency of (34)-(36) with the result of
classical theory, (5) and (6); in particular, (35) signifies that
the region of the classical shadow is not accessible to a
quasiclassical particle. On the contrary, the region of the
classical double image exhibits itself in a way which differs from
that in classical theory, compare (37) with the classical
differential cross section,
\begin{equation}
\frac{{\rm d}\sigma^{({\rm class})}}{{\rm d}\varphi}=r_c(1-\eta)^2 \cos\left[\frac 12(1-\eta)\varphi\right]
\sin\left(\frac 12\eta\pi\right),\quad \varphi_\eta<\varphi<-\varphi_\eta, \label{eq38}
\end{equation}
which is the sum of (5) and (6). In particular, we get in the
strictly forward direction
\begin{equation}
\left.\frac{{\rm d}\sigma^{({\rm q-class})}}{{\rm d}\varphi}\right|_{\varphi=0}=
2\left.\frac{{\rm d}\sigma^{({\rm class})}}{{\rm d}\varphi}\right|_{\varphi=0}=
2r_c(1-\eta)^2\sin\left(\frac 12\eta\pi\right).\label{eq39}
\end{equation}
However, integrating either (34) and (35), or (36) and (37) over the
whole range of the scattering angle
$(-|\varphi_\eta|<\varphi<2\pi-|\varphi_\eta|)$ \footnote{ Note that
the second term in figure brackets in (37) yields, upon integration,
the contribution which is of order $r_c O[(kr_c)^{-3/2}]$, and
thereby the first term contributes only.}, one gets
\begin{equation}
\sigma^{({\rm q-class})}=2r_c(1-\eta), \label{eq40}
\end{equation}
which coincides with the total cross section in classical theory,
see (7) and (10).

As to the Fraunh\"{o}fer diffraction, the contribution of the
interference between the diffraction peaks, $f_+^{({\rm
peak})}\left(f_-^{({\rm peak})}\right)^*+f_-^{({\rm
peak})}\left(f_+^{({\rm peak})}\right)^*$, is negligible comparing
to the contribution of each peak,
\begin{equation}
\frac{{\rm d}\sigma^{(\rm peak)}_\pm}{{\rm d}\varphi}=|f_\pm^{({\rm peak})}(k,\,\varphi\pm\varphi_\eta)|^2=
r_c(1-\eta)\Delta_{\frac 12kr_c(1-\eta)}(\varphi\pm\varphi_\eta). \label{eq41}
\end{equation}
The appropriate integrated cross section is
\begin{equation}
\sigma^{({\rm peak})}_\pm=r_c(1-\eta). \label{eq42}
\end{equation}
and, thus, the total cross section,
\begin{equation}
\sigma_{\rm tot}=\sigma^{({\rm q-class})}+\sigma^{(\rm peak)}_++\sigma^{(\rm peak)}_-=4r_c(1-\eta), \label{eq43}
\end{equation}
is twice the classical total cross section.

\section{Quasiclassical limit of quantum-mechanical scattering off an impenetrable magnetic vortex in conical space}

We consider scattering off an impenetrable tube which is filled with
magnetic field with total flux $\Phi$. The quantum-mechanical wave
function obeys condition (12) at the edge of the tube, and the
Schr\"{o}dinger hamiltonian outside the tube takes form
\begin{equation}
{\rm H}=-\frac{\hbar^2}{2m}\left[\partial_r^2+\frac 1r\partial_r+\frac{1}{(1-\eta)^2r^2}
\left(\partial_\varphi-{\rm i\frac{\Phi}{\Phi_0}}\right)^2+\partial_z^2\right], \label{eq44}
\end{equation}
where $\Phi_0=2\pi\hbar{\rm c}{\rm e}^{-1}$ is the London flux
quantum. The exact expression for the scattering amplitude in this
case is
\begin{eqnarray}
f(k,\,\varphi)=-\exp[2{\rm i}k(r_{\rm c}-\xi_{\rm c})]\sqrt{\frac{2}{\pi {\rm i}k}}\sum\limits_{n\in\mathbb{Z}}
e^{{\rm in}(\varphi-\pi)}\left[\frac {\rm i}2\sin(\alpha_n\pi)+\right.\nonumber \\
\left.+\exp(-{\rm i}\alpha_n\pi)\frac{J_{\alpha_n}(kr_c)}{{\rm H}^{(1)}_{\alpha_n}(kr_c)}\right],\label{eq45}
\end{eqnarray}
where
\begin{equation}
\alpha_n=|n-\Phi/\Phi_0|(1-\eta)^{-1}.\label{eq46}
\end{equation}
The contribution of the infinite sum with the sine term in square
brackets in (45) yields the amplitude for scattering off a vortex of
zero transverse size, see (A.6) in Appendix, which is negligible in
the quasiclassical limit. The analysis of the remaining infinite sum
which contains the quotient of the cylindrical functions yields that
the dominant contribution is given by terms with $\alpha_n \leq
kr_c$ in the case $kr_c\gg 1$, and the scattering amplitude in this
case takes form (29), where the contribution of the Fraunh\"{o}fer
diffraction on the vortex is given by
\begin{eqnarray}
f_\pm^{({\rm peak})}(k,\,\varphi\pm \varphi_\eta)=
-\frac{\exp\left[2{\rm i}k(r_{\rm c}-\xi_{\rm c})\right]}{\sqrt{2\pi ik}}\,
\frac{\sin\left[\frac 12kr_c(1-\eta)(\varphi\pm\varphi_\eta)\right]}
{\sin\left[\frac 12(\varphi\pm\varphi_\eta)\right]}\times
\nonumber \\
\times\exp\left\{\pm{\rm i}\Phi\Phi_0^{-1}(\pi-\varphi_\eta)+{\rm i}\left[[\![\Phi\Phi_0^{-1}]\!]+\frac
12\pm\frac 12kr_c(1-\eta)\right](\varphi\pm\varphi_\eta)\right\},\label{eq47}
\end{eqnarray}
and the contribution of the quasiclassical reflection from the
vortex is given by (see \cite{Si5})
\begin{eqnarray}
f^{(\rm q-class)}(k,\,\varphi)=-\exp[2{\rm i}k(r_{\rm c}-\xi_{\rm c})](1-\eta)\sqrt{\frac{r_{\rm c}}{2}}
\sum\limits_{l}\sqrt{\cos\left[\frac 12(1-\eta)(\varphi-\pi+2l\pi)\right]}\times \nonumber \\
\times\exp\left\{{\rm i}\Phi\Phi_0^{-1}
(\varphi-\pi+2l\pi)-2{\rm i}kr_{\rm c}\cos\left[\frac 12(1-\eta)(\varphi-\pi+2l\pi)\right]\right\},\label{eq48}
\end{eqnarray}
with integers $l$ satisfying condition (32). It should be noted that
(48) is valid for $-\infty<\eta<1$, as well as is the following
expression for the incident wave in this case,
\begin{equation}
\psi_{\rm in}({\bf r};\,{\bf k})=(1-\eta)\sum\limits_{l}\exp\{-{\rm i}kr\cos[(1-\eta)(\varphi
-\pi+2l\pi)]\}\exp[{\rm i}\Phi\Phi_0^{-1}(\varphi-\pi+2l\pi)],\label{eq49}
\end{equation}
where the sumation is over the same integers $l$.

Thus, the crucial point is, whether the sum over $l$ contains more
than one term. If this is the case, then, due to the interference of
different terms, the differential cross section corresponding to
(48) is dependent on the value of the vortex flux.

In the case $-\infty<\eta\leq 0$ the differential cross section
corresponding to (48) is given by (34) and (35), and this describes
classical scattering which is independent of $\Phi$. The same is
true for the differential cross section out of the double-image
region in the case $0<\eta<1/2$, see (36). On the contrary, in the
double-image region in the case $0<\eta<1/2$, the differential cross
section corresponding to (48) takes the form which differs both from
(37) and (38):
\begin{eqnarray}
\frac{{\rm d}\sigma^{({\rm q-class})}}{{\rm d}\varphi}=r_{\rm c}(1-\eta)^2\Biggl\{\cos\left[\frac 12(1-\eta)\varphi\right]
\sin\left(\frac 12\eta\pi\right)+ \Biggr.\nonumber \\
\Biggl.+\sqrt{\sin^2\!\left(\frac 12\eta\pi\right)\!-\!\sin^2\!\left(\frac 12(1\!-\!\eta)\varphi\right)}
\cos\!\left[2\Phi\Phi_0^{-1}\pi\!+\!4kr_{\rm c}\sin\!\left(\frac 12(1\!-\!\eta)\varphi\right)\cos\!\left(
\frac 12\eta\pi\right)\right]\Biggr\}, \nonumber \\
\varphi_\eta<\varphi<-\varphi_\eta.\label{eq50}
\end{eqnarray}
In the forward direction in the case
$\sin\left(\frac{1}{2kr_c}\right)\ll\sin\left(\frac12\eta\pi\right)$,
we get
\begin{eqnarray}
\frac{{\rm d}\sigma^{({\rm q-class})}}{{\rm d}\varphi}=2r_{\rm c}(1-\eta)^2\sin(\frac 12\eta\pi)\cos^2
\left[\Phi\Phi_0^{-1}\pi+ kr_c(1-\eta)\varphi\cos\left(\frac 12\eta\pi\right)\right],
\nonumber \\-(kr_c)^{-1}<(1-\eta)\varphi<(kr_c)^{-1}.\label{eq51}
\end{eqnarray}
In particular, in the strictly forward direction, $\varphi=0$, we
get (39) at $\Phi=n\Phi_0$ and zero at $\Phi=\left(n+\frac
12\right)\Phi_0$ \,\,\,($n\in \mathbb{Z}$).

The vanishing of the quasiclassical differential cross section in
the strictly forward direction, when the vortex flux equals a
half-of-odd-integer multiple of the London flux quantum, is a
consequence of a more general result which is valid for all
wavelengths of a scattered particle. Using the exact expression for
the scattering wave solution to the Schr\"{o}dinger equation, see
(A.7) in Appendix, one can get
\begin{eqnarray}
\left.\psi({\bf r};\,{\bf k})\right|_{\Phi=(n+\frac12)\Phi_0}=2{\rm i}
\exp[{\rm i}k(r_c-\xi_c)+{\rm i}\varphi/2]\sqrt{\frac{r}{r-r_c+\xi_c}}\times
\nonumber \\ \times \sum\limits_{n'\geq 1}\sin\left[\left(n'-\frac 12\right)\varphi\right]
\exp\left[{\rm i}n'\pi-\frac{{\rm i}(n'-1/2)\pi}{2(1-\eta)}\right]\left[J_{(n'-1/2)(1-\eta)^{-1}}(kr)\right.-\nonumber \\
\left.-\frac{J_{(n'-1/2)(1-\eta)^{-1}}(kr_c)}{H^{(1)}_{(n'-1/2)(1-\eta)^{-1}}(kr_c)}
H^{(1)}_{(n'-1/2)(1-\eta)^{-1}}(kr)\right], \label{eq52}
\end{eqnarray}
and, thus, wave function (52) vanishes at $\varphi=0$ for all values
of $k$.

The dependence of the differential cross section on the vortex flux
is washed off after integration over the whole range of the
scattering angle, and we get cross section (40) which is independent
of the vortex flux and coincides with the classical total cross
section, see (10).

As to the Fraunh\"{o}fer diffraction, the differential cross section
corresponding to either of two diffraction peaks is given by (41),
being independent of the vortex flux. The appropriate integrated
cross section is given by (42), and, hence, the total cross section
is given by (43).

These results should be compared with the results for the
quasiclassical limit of quantum-mechanical scattering off an
impenetrable magnetic vortex in Euclidean space ($\eta=0$). The
scattering amplitude in the latter case takes form
\begin{equation}
f(k,\,\varphi)=f^{({\rm peak})}(k,\,\varphi)+
f^{({\rm class})}(k,\,\varphi)+\sqrt{r_c}O[(kr_c)^{-1/6}],\label{eq53}
\end{equation}
where the contribution of the reflection from the vortex is given by
$f^{(\rm class)}$ (23), and the contribution of the Fraunh\"{o}fer
diffraction on the vortex is given by
\begin{equation}
f^{({\rm peak})}(k,\,\varphi)=-\sqrt{\frac{2}{\pi{\rm i}k}}\frac{\sin\left(\frac 12kr_c\varphi\right)}{\sin\left(
\varphi/2\right)}\cos\left(\frac 12kr_c\varphi+\Phi\Phi_0^{-1}\pi\right)
\exp\left[{\rm i}\left([\![\Phi\Phi_0^{-1}]\!]+\frac 12\right)\varphi\right].\label{eq54}
\end{equation}
Thus, the reflection is consistent with classical theory, see (24),
whereas the forward diffraction peak is dependent on the vortex
flux, cf. (21):
\begin{eqnarray}
\frac{{\rm d}\sigma^{({\rm peak})}}{{\rm d}\varphi}=2r_c\left\{\cos(2\Phi\,\Phi_0^{-1}\pi)
\Delta_{kr_c}(\varphi)+\right. \nonumber \\ \left.+\left[1-\cos(2\Phi\,\Phi_0^{-1}\pi)
-\sin(2\Phi\,\Phi_0^{-1}\pi)\sin(kr_c\varphi)\right]\Delta_{\frac 12kr_c}(\varphi)\right\}.\label{eq55}
\end{eqnarray}
Explicitly, in the forward direction we get

\begin{eqnarray}
\frac{{\rm d}\sigma^{({\rm peak})}}{{\rm d}\varphi}=\frac{2}{\pi}kr_c^2\left\{\cos^2\left(\Phi\Phi_0^{-1}\pi\right)
-\frac 12kr_c\varphi \sin\left(2\Phi\Phi_0^{-1}\pi\right)-\right.  \nonumber \\ \left.-\frac{1}{24}(kr_c\varphi)^2\left[
1+7\cos(2\Phi\Phi_0^{-1}\pi)\right]\right\}\left\{1+O[(kr_c\varphi)^2]
\right\},\,\,\,\,\,\, |\varphi|\ll(kr_c)^{-1}.\label{eq56}
\end{eqnarray}
The integrated cross section, $\sigma^{{(\rm peak)}}$, is
independent of the vortex flux and is equal to (22). The total cross
section is equal to (26) that is twice the classical total cross
section.

\section{Results and discussion}

In the present paper we have considered quantum-mechanical
scattering of a nonrelativistic short-wavelength particle by an
impenetrable magnetic vortex of the finite transverse size. If space
outside the vortex is Euclidean, then scattering may be decomposed
into the classical reflection according to the laws of geometric
optics, see (24), which is smoothly distributed in all directions,
and the Fraunh\"{o}fer diffraction which is strongly peaked in the
forward direction, see (55); the latter depends on the vortex flux
with the period equal to the London flux quantum, whereas the former
does not.

This picture is changed in the case when space outside the vortex is
conical. If $-\infty<\eta <1/2$, then the Fraunh\"{o}fer diffraction
is peaked in two directions, see (41) where $\varphi_\eta$ is
defined by (3), and is independent of the vortex flux. If $\eta<0$,
then the forward region between the diffraction peaks is the region
of the classical shadow and is inaccessible to a quasiclassical
particle, see (35); scattering outside the shadow region occurs as
the classical reflection, see (34). If $0<\eta<1/2$, then the
forward region between the diffraction peaks is the region of the
classical double image. Although scattering outside the double-image
region occurs as the classical reflection, see (36), scattering
inside the double-image region differs from that in classical theory
and is dependent on the vortex flux with the period equal to the
London flux quantum, see (50). Hence, the Aharonov-Bohm effect
persists in the quasiclassical limit for scattering angles in the
range of the double-image region in conical space with $0<\eta<1/2$.
The physical reason of this persistence is the Fresnel diffraction
that is the diffraction in converging rays (whereas the
Fraunh\"{o}fer diffraction is the diffraction in almost parallel
rays).

Thus, we can summarize that the scattering Aharonov-Bohm effect
persists in the quasiclassical limit owing to the diffraction.
Although the effect is invisible for the cross section integrated
over the whole range of the scattering angle, the effect reveals
itself for the differential cross section, i.e. for the product of
the distance to the scatterer and the intensity of the scattered
wave in the units of the intensity of the incident wave. In the case
of a magnetic vortex in Euclidean space, the persistence of the
Aharonov-Bohm effect is due to the Fraunh\"{o}fer diffraction which
is peaked in the forward direction. In the case of a magnetic vortex
in conical space, the peak of the Fraunh\"{o}fer diffraction is
shifted from the forward direction and splitted into two peaks in
directions which are symmetric with respect to the forward one; the
contribution of each peak is independent of the vortex flux. If the
forward region between two Fraunh\"{o}fer-diffraction peaks is the
region of the classical shadow, then the Aharonov-Bohm effect
disappears in the quasiclassical limit. If the forward region
between two Fraunh\"{o}fer-diffraction peaks is the region of the
classical double image, then the persistence of the Aharonov-Bohm
effect in the quasiclassical limit is due to the Fresnel diffraction
in this region.

Since a peak of the Fraunh\"{o}fer diffraction is elusive to
experimental measurements, it might be hard to detect the vortex
flux dependence in the strictly forward direction in Euclidean
space. On the contrary, the vortex flux dependence which is due to
the Fresnel diffraction in conical space looks much more likely to
be detectable: it is spread over the wider region in the forward
direction and its amount is finite in the quasiclassical limit,
compare (51) with (56). It should be noted that the optical theorem
in conical space imposes no restrictions on the scattering amplitude
in the strictly forward direction; instead, it involves the
scattering amplitude in two strict directions of the
Fraunh\"{o}fer-diffraction peaks, see (A.9) in Appendix.

Another distinction of scattering in conical space is that it
depends on spin of a scattered particle: the appropriate spin
connection which is dependent on $\eta$ should be introduced in
hamiltonian (44). In particular, for a spin-1/2 particle the results
are modified in the following way (see \cite{Si5}): one should
change $\Phi\Phi_0^{-1}$ to $\Phi\Phi_0^{-1}\mp \frac 12\eta$, where
two signs correspond to two spin states which are defined by
projections of spin on the vortex axis.

If a magnetic vortex is trapped inside a superconducting shell
\footnote{ Since no magnetic field can leak outside, the
superconducting shell guarantees for certain that there is no
overlap between the region of magnetic flux and the region which is
accessible to the scattered particle, see \cite{Pes,Ton}.}, then its
flux is quantized in the units of a semifluxon, i.e. half of the
London flux quantum. In view of (51) we get the following relation
in the quasiclassical limit when condition
$\sin\left(\frac{1}{2kr_c}\right)\ll\sin\left(\frac
12\eta\pi\right)$ is satisfied:
\begin{equation}
\left.\frac{{\rm d}\sigma^{({\rm q-class})}}{{\rm d}\varphi}\right|_{\Phi=n\Phi_0}+
\left.\frac{{\rm d}\sigma^{({\rm q-class})}}{{\rm d}\varphi}\right|_{\Phi=
\left(n+\frac 12\right)\Phi_0}=2\left.\frac{{\rm d}\sigma^{({\rm class})}}{{\rm d}\varphi}\right|_{\varphi=0},\,\,\,\,
(1-\eta)|\varphi|<(kr_c)^{-1}, \label{eq57}
\end{equation}
where the classical differential cross section is given by (38).
Hence, the quasiclassical limit of the Aharonov-Bohm effect in the
case when the vortex flux equals an integer multiple of a semifluxon
can be presented in the form
\begin{equation}
\left.\frac{{\rm d}\sigma^{({\rm q-class})}}{{\rm d}\varphi}\right|_{\Phi=n\Phi_0/2}=\left.F(\varphi,\,\pm)
\frac{d\sigma^{({\rm class})}}{d\varphi}\right|_{\varphi=0},\,\,\,\,
(1-\eta)|\varphi|<(kr_c)^{-1}, \label{eq58}
\end{equation}
where the upper (lower) sign in $F$ corresponds to even (odd) $n$,
and
\begin{equation}
F(\varphi,\,\pm)=1\pm\cos\left[2kr_c(1-\eta)\varphi\cos\left(\frac 12\eta\pi\right)\right] \label{eq59}
\end{equation}
for a spinless particle,
\begin{equation}
F(\varphi,\,\pm)=1\pm\cos\left[2kr_c(1-\eta)\varphi\cos\left(\frac 12\eta\pi\right)\right]\cos(\eta\pi) \label{eq60}
\end{equation}
for an unpolarized spin-1/2 particle,
\begin{eqnarray}
F(\varphi,\,\pm)=1\pm\cos\left[2kr_c(1-\eta)\varphi\cos\left(\frac 12\eta\pi\right)\right]\cos(\eta\pi)\pm\nonumber \\
\pm\sigma\sin\left[2kr_c(1-\eta)\varphi\cos\left(\frac 12\eta\pi\right)\right]\sin(\eta\pi) \label{eq61}
\end{eqnarray}
for a polarized spin-1/2 particle ($\sigma=\pm1$ correspond to two
polarization states).

\section{Conclusions}

Although the Aharonov-Bohm effect is the purely quantum effect that
is alien to classical physics, it persists in the quasiclassical
limit owing to the diffraction persisting in the short-wavelength
limit in the region of forward scattering. To be more precise, the
persistence of the scattering Aharonov-Bohm effect for
quasiclassical nonrelativistic particles is due to the
Fraunh\"{o}fer diffraction in the case when space outside the
enclosed magnetic flux is Euclidean, and the Fresnel diffraction in
the case when the outer space is conical. Hence, the enclosed
magnetic flux serves as a gate for the propagation of
short-wavelength, almost classical, particles. In the case of
conical space, this quasiclassical effect which is in principle
detectable depends on the particle spin, and the most efficient gate
is for the propagation of spinless particles in the strictly forward
direction, see (59) at $\varphi=0$. For instance, the propagation of
fast-moving electronic excitations in the bilayer graphene sample
\footnote{ The relevance of hamiltonian (44) rather than the Dirac
hamiltonian is due to the quadratic dispersion law in this case, see
\cite{gra}.} of conical shape can be governed by the magnetic flux
applied through a hole at the tip.

\section*{Acknowledgements}

We would like to thank R. Jackiw for helpful comments. The work was
partially supported by the Department of Physics and Astronomy of
the National Academy of Sciences of Ukraine under special program
``Fundamental properties of physical systems in extremal
conditions''.

\renewcommand{\thesection}{A}
\renewcommand{\theequation}{\thesection.\arabic{equation}}
\setcounter{section}{1} \setcounter{equation}{0}

\section*{Appendix: Solution to the scattering problem and $S$-matrix}

In the problem of quantum-mechanical scattering off a vortex in
conical space one encounters a situation when the interaction with a
scattering centre (vortex) is not of the potential type and is even
nondecreasing at large distances from the centre. Really, the
interaction hamiltonian corresponding to (44) can be presented as
\begin{equation}\label{a1}
    \biggl.H-H\biggr|_{ \Phi=0\atop \eta=0 }
     =v({\bf r})+v^j({\bf r})\left(-{\rm i}\frac{\partial}{\partial r^j}\right)+v^{jj'}({\bf r})
     \left(-\frac{\partial^2}{\partial r^j\partial r^{j'}}\right),
\end{equation}
where $r^1=r\cos \varphi$, $r^2=r\sin \varphi$\,\,\,
$(j,\,j'=1,\,2)$, and
\begin{equation}\label{a2}
v\sim O(r^{-2}),\,\,\,\, v^j\sim O(r^{-1})\,\,\,\,{\rm and}\,\,\,\, v^{jj'}\sim O(1),\,\,\,\,r\rightarrow \infty;
\end{equation}
note that $v^j$, in contrast to $v$ and $v^{jj'}$, is a complex
function (to be more precise, the imaginary part of $v^j$ of order
$r^{-1}$ is due to nondecrease of real quantity $v^{jj'}$ in the
limit $r\rightarrow\infty$). Methods of the standard scattering
theory (see, e.g., \cite{Ree}) can be implemented in the case of the
short-range interaction, i.e. when coefficient functions $v$, $v^j$,
and $v^{jj'}$ decrease faster than $O(r^{-1-\varepsilon})$ in the
limit $r\rightarrow \infty$ ($\varepsilon>0$). The uttermost
advancement to the case of the long-range interaction is due, to our
knowledge, to H\"{o}rmander \cite{Hor} who considered a class of
interactions with a long-range part which is characterized by real
coefficient functions decreasing as $O(r^{-\varepsilon})$
($0<\varepsilon\leq 1$). As he notes in his monograph \cite{Hor},
"the existence of modified wave operators is proved under the
weakest sufficient conditions among all those known at the present
time".

H\"{o}rmander's conditions are satisfied by the interaction in the
problem of scattering off a vortex in Euclidean space ($\Phi\neq 0$
and $\eta=0$),
$$
v\sim O(r^{-2}) \quad {\rm and}\quad v^j\sim O(r^{-1}),\quad r\rightarrow\infty
$$
($v$ and $v^j$ are real, and $v^{jj'}=0$), and, for instance, by the
interaction in the problem of scattering off a Coulomb centre,
$$
v\sim O(r^{-1}),\quad r\rightarrow\infty
$$
($v$ is real, and $v^j=v^{jj'}=0$). On the contrary, H\"{o}rmander's
conditions are violated by the interaction in the problem of
scattering off a vortex in conical space ($\Phi\neq 0$ and $\eta\neq
0$). Nevertheless, scattering theory can be constructed even in the
last case, and this has been done in \cite{Si5}, basing on earlier
works \cite{Ho,Des,Sou,Si2}.

In this framework involving the use of both the time-dependent
scattering theory and the Lippmann-Schwinger equation, one gets the
following expression for the $S$-matrix in the case of scattering of
a nonrelativistic particle off an impenetrable vortex of transverse
radius $r_c$:
\begin{eqnarray}\label{a3}
S(k,\,\varphi,\,k_z;\,k',\,\varphi',\,k'_z)=\frac{\delta(k-k')}{\sqrt{kk'}}\,\frac{e^{2{\rm i}k(r_c-\xi_c)}}{2\pi}
\sum\limits_{n\in\mathbb{Z}}e^{{\rm i}
n(\varphi-\varphi'-\pi)-{\rm i}\alpha_n\pi}\times\nonumber \\
\times\left[1-2\frac{J_{\alpha_n}(kr_c)}{H^{(1)}_{\alpha_n}(kr_c)}\right] \delta(k_z-k'_z),
\end{eqnarray}
where the final $({\bf k})$ and initial $({\bf k}')$ two-dimensional
wave vectors of the particle are written in polar variables, and
$\alpha_n$ is given by (46). The $S$-matrix is decomposed as
\begin{equation}\label{a4}
S(k,\,\varphi,\,k_z;\,k',\,\varphi',\,k'_z)=\left[\tilde{I}(k,\,\varphi;\,k',\,\varphi')+
\frac{\delta(k-k')}{\sqrt{kk'}}\sqrt{\frac{{\rm i}k}{2\pi}}f(k,\,\varphi-\varphi')\right]\delta(k_z-k'_z),
\end{equation}
where $f(k,\,\varphi)$ is given by (45) and
\begin{eqnarray}\label{a5}
\tilde{I}(k,\,\varphi;\,k',\,\varphi')=\frac{\delta(k-k')}{\sqrt{kk'}}\,\frac{e^{2{\rm i}k(r_c-\xi_c)}}{2\pi}
\sum\limits_{n\in\mathbb{Z}}e^{{\rm i}n(\varphi-\varphi'-\pi)}\cos(\alpha_n\pi)= \nonumber \\
=\frac 12\frac{\delta(k-k')}{\sqrt{kk'}}e^{2{\rm i}k(r_c-\xi_c)}\left\{\exp
\left[{\rm i}\Phi\Phi_0^{-1}(\pi-\varphi_\eta)\right]\Delta(\varphi-\varphi'+\varphi_\eta)+\right. \nonumber \\
\left.+\exp\left[-{\rm i}\Phi\Phi_0^{-1}(\pi-\varphi_\eta)\right]\Delta(\varphi-\varphi'-\varphi_\eta)\right\}.
\end{eqnarray}
Here, all delta-functions of the angular variables, stemming from
the infinite sum in (A.3), are gathered in the modified unity matrix
$\tilde{I}$, whereas scattering amplitude $f$ is free of
delta-functions. Thus the long-range nature of interaction manifests
itself in the distortion of the usual unity matrix,
$I(k,\,\varphi;\,k',\,\varphi')=\frac{\delta(k-k')}{\sqrt{kk'}}\Delta(\varphi-\varphi')$,
the latter being appropriate to the case of the short-range
interaction (for instance, in the problem of scattering off an
impenetrable tube in Euclidean space). The modified unity matrix
contains angular delta-functions which are peaked in two directions
of the Fraunh\"{o}fer diffraction,
$\varphi-\varphi'=\mp\varphi_\eta$. Due to this circumstance, the
incident wave is also distorted, taking the form of (49).

If one neglects the transverse size of the vortex by putting
$r_c=0$, then the contribution of the cylindrical functions in (45)
vanishes and the scattering amplitude takes form
\begin{eqnarray}\label{a6}
&&f(k,\,\varphi)=-\sqrt{\frac{\rm i}{2\pi k}}\sum\limits_{n\in\mathbb{Z}}e^{{\rm i}n(\varphi-\pi)}\sin(\alpha_n\pi)=\nonumber \\
&&=\frac 12\sqrt{\frac{{\rm i}}{2\pi k}}\left\{\exp\left[{\rm i}\Phi\Phi_0^{-1}(\pi-\varphi_\eta)+
{\rm i}[\![\Phi\Phi_0^{-1}]\!](\varphi+\varphi_\eta)\right]\left[\cot\left(\frac 12(\varphi+\varphi_\eta)\right)+
{\rm i}\right]-\right. \nonumber \\
&&\left.-\exp\left[-{\rm i}\Phi\Phi_0^{-1}(\pi-\varphi_\eta)+{\rm i}[\![\Phi\Phi_0^{-1}]\!](\varphi-\varphi_\eta)\right]
\left[\cot\left(\frac 12\left(\varphi-\varphi_\eta\right)\right)+{\rm i}\right]\right\}.
\end{eqnarray}
This expression is appropriate to the ultraquantum limit, i.e. to
the limit of long wavelengths of a scattered particle, $k\rightarrow
0$. The amplitude in this limit diverges in two directions
corresponding to the peaks of the Fraunh\"{o}fer diffraction, and,
therefore, the total cross section in this limit diverges as well.
However, this divergence, as well as the discontinuities of the
incident wave (49) at the positions of the diffraction peaks, is an
artefact of the expansion in powers of $r^{-1/2}$ in (11), which is
invalid at the positions of the diffraction peaks. Really, using the
exact expression for the whole wave function
\begin{eqnarray}\label{a7}
\psi({\bf r};\,{\bf k})=e^{{\rm i}k(r_c-\xi_c)}\sqrt{\frac{r}{r-r_c+\xi_c}}\sum\limits_{n\in \mathbb{Z}}
e^{{\rm i}n(\varphi-\pi)-\frac {\rm i}{2}\alpha_n\pi}\times \nonumber \\
\times \left[J_{\alpha_n}(kr)-\frac{J_{\alpha_n}(kr_c)}{H_{\alpha_n}^{(1)}(kr_c)}H_{\alpha_n}^{(1)}(kr)\right],
\end{eqnarray}
one gets, in particular, in the case $0<\eta<1/2$ and $kr\gg 1$,
$kr_c\ll 1$ (for a more general case see \cite{Si5}):
\begin{eqnarray}\label{a8}
&&\psi({\bf r};\,{\bf k})=\frac 12(1-\eta)\exp\left[{\rm i}kr\pm {\rm i}\Phi\Phi_0^{-1}(\pi-\varphi_\eta)\right]
+O\left[(kr)^{-1/2}\right]+ \nonumber \\
&&+(1-\eta)\exp\left[{\rm i}kr\cos(2\eta\pi)\mp{\rm i}\Phi\Phi_0^{-1}(\pi+\varphi_\eta)\right]
\left[1\mp{\rm i}kr(1-\eta)(\varphi\pm\varphi_\eta)\sin(2\eta\pi)\right]\times \nonumber \\
&&\times\left\{1+O\left[(kr)^{-1/2}\right]\right\}+O\left\{\left[kr(\varphi\pm\varphi_\eta)\right]^2\right\},\,\,\,\,
kr|\varphi\pm \varphi_\eta|\ll 1,
\end{eqnarray}
which is finite and continuous at $\varphi=\mp\varphi_\eta$.

One more manifestation of the long-range nature of interaction is
that the optical theorem in the quasiclassical limit involves
regularized angular delta-function (19):
\begin{eqnarray}\label{a9}
&\cos\left[\Phi\Phi_0^{-1}(\pi-\varphi_\eta)\right]\sqrt{\frac{2\pi}{k}}\,{\rm Im}
\left\{{\rm i}^{-1/2}e^{-2{\rm i}k(r_c-\xi_c)}\left[f_+^{({\rm peak})}(k,\,0)+f_-^{({\rm peak})}(k,\,0)\right]\right\} -
\nonumber \\
&-\sin\left[\Phi\Phi_0^{-1}(\pi-\varphi_\eta)\right]\sqrt{\frac{2\pi}{k}}\,{\rm Re}\left\{
{\rm i}^{-1/2}e^{-2{\rm i}k(r_c-\xi_c)}\left[f_+^{({\rm peak})}
(k,\,0)-f_-^{({\rm peak})}(k,\,0)\right]\right\}+ \nonumber \\
&+\frac{2\pi}{k}\Delta_{kr_c(1-\eta)}(0)=\sigma_{\rm tot}.
\end{eqnarray}
In the case of a vortex in Euclidean space ($\eta=0$), the optical
theorem in the quasiclassical limit takes form
\begin{equation}\label{a10}
2\cos(\Phi\Phi_0^{-1}\pi)\sqrt{\frac{2\pi}{k}}
\,{\rm Im}\left[{\rm i}^{-1/2}f^{({\rm peak})}(k,\,0)\right]+
\sin^2(\Phi\Phi_0^{-1}\pi)\frac{4\pi}{k}\Delta_{kr_c}(0)= \sigma_{\rm tot}.
\end{equation}

\newpage

\begin{figure}
\includegraphics[width=135mm]{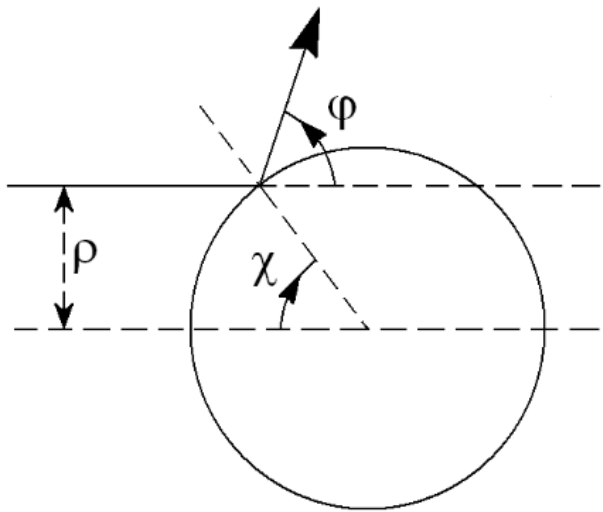}
\caption{ Scattering on a core with the incidence angle equal to the reflection angle.}
\end{figure}

\newpage

\begin{figure}
\includegraphics[width=140mm]{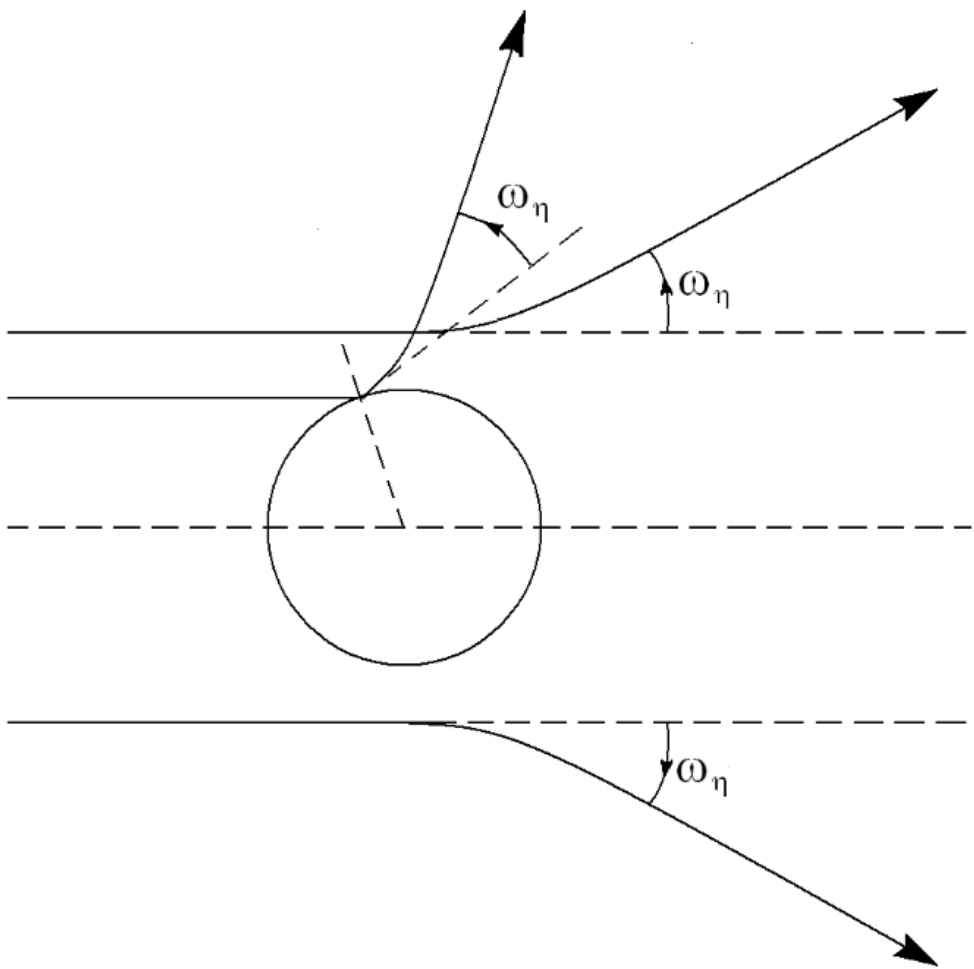}
\caption{ Scattering in conical space with $-\infty<\eta<0$: $\omega_\eta=-\eta\pi(1-\eta)^{-1}$.}
\end{figure}

\newpage

\begin{figure}
\includegraphics[width=145mm]{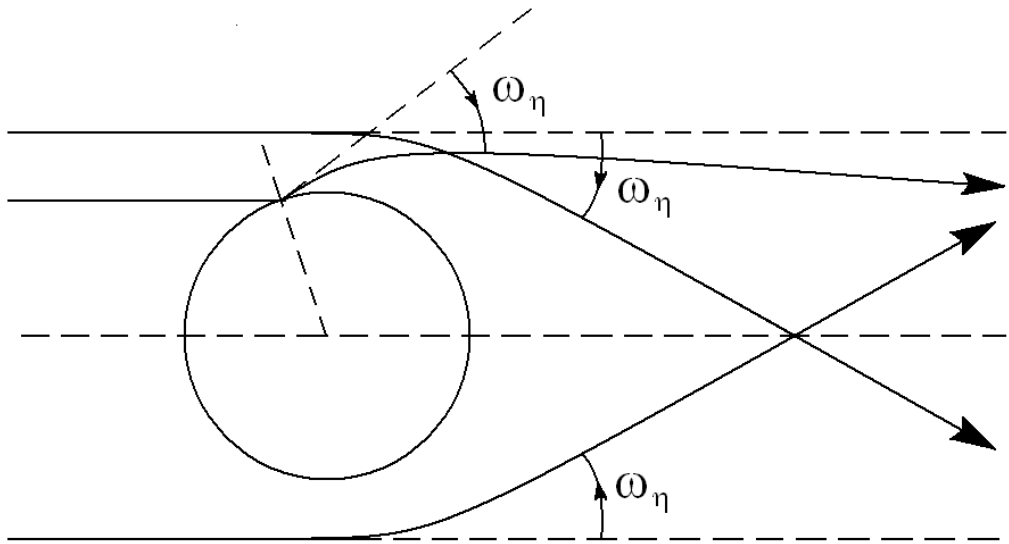}
\caption{ Scattering in conical space with $0<\eta<1/2$: $\omega_\eta=\eta\pi(1-\eta)^{-1}$.}
\end{figure}

\end{document}